\documentclass[aps,twocolumn,
showpacs,prb]{revtex4}
\usepackage{graphicx}
\usepackage{amsmath}
\usepackage{upgreek}

\begin{document}
\title{Transport properties controlled by a thermostat: An extended dissipative particle dynamics thermostat}
\author{Christoph Junghans}
\author{Matej Praprotnik}
\altaffiliation{On leave from the National Institute of Chemistry, Hajdrihova 19,
                 SI-1001 Ljubljana, Slovenia.}
\author{Kurt Kremer}
\affiliation{Max-Planck-Institut f\"ur Polymerforschung, Ackermannweg 10, D-55128 Mainz, Germany}
\begin{abstract}
We introduce a variation of the dissipative particle dynamics (DPD)
thermostat that allows for controlling transport properties of molecular
fluids. The standard DPD thermostat acts only on a relative velocity along the interatomic axis.
Our extension includes the damping of the perpendicular components of
the relative velocity,  yet keeping the advantages
of conserving Galilei invariance and within our error bar also hydrodynamics.
This leads to a second friction parameter for tuning the transport properties of the system.
Numerical simulations of a simple Lennard-Jones fluid and liquid
water demonstrate a~very sensitive behaviour
of the transport properties, e.g., viscosity, on the strength of the
new friction parameter. 
We envisage that the new thermostat will be 
very useful for the coarse-grained and adaptive resolution simulations of soft
matter, where the diffusion constants and viscosity of the
coarse-grained models are typically too
high/low, respectively, compared to all-atom simulations.
\end{abstract}
\maketitle
\section{Introduction}

Using the optimal set of degrees
of freedom (DOFs) in computer simulations of soft matter 
guarantees efficiency, accuracy and avoids huge
amounts of unnecessary detail, which might even obscure the
underlying physics. This very idea is exploited in systematic coarse-graining efforts
for modern computational materials science and biophysics
problems~\cite{Harmandaris:2006,Hess:2006,Nico:2007},
where atomistic simulations are usually beyond the
possibilities of current and near future computers. 
A similar philosophy of reducing the number of DOFs is also employed 
by representing clusters of molecules with soft particles when
simulating fluids on a mesoscopic scale using dissipative particle dynamics
(DPD)\cite{Hoogerbrugge:1992, Koelman:1993, Espanol:1995,
  Groot:1997,Espanol:1998}. 
However, in typical soft matter systems different time- and
length-scales are intrinsically interconnected and a
multiscale modeling approach is required to tackle such
problems in the most efficient way\cite{DelleSite:2002,Delgado:2003,Barsky:2004,Delgado:2005,Neri:2005,Villa:2005,Fabritiis:2006,Voth2006}.

Recently, we have proposed an adaptive resolution scheme (AdResS) that
couples the atomistic and coarse-grained levels of detail\cite{Praprotnik:2005:4,Praprotnik:2006,
Praprotnik:2007,Praprotnik:2007:1,Praprotnik:2007:3,Praprotnik:2007:2}.
Due to the reduction in DOFs upon coarse-graining, which eliminates the
fluctuating forces associated with those missing molecular DOFs, the
coarse-grained molecules typically move faster than the corresponding
atomistically resolved ones\cite{Tschop:1998,Izvekov:2006}.
 Although the accelerated dynamics is advantageous in some cases it can turn out to be 
problematic if one is really interested in dynamics in situations where two levels of
resolutions are used within one simulation, as in the case of AdResS. 
To overcome this problem
one can couple different classes of DOFs to the Langevin
thermostat with different friction constants\cite{carmeli,moix,grigolini,voth}.
We have shown in the example of liquid water that 
the coarse-grained dynamics can be slowed down 
by increasing the effective friction in the coarse-grained
system\cite{water2}. However, it is well known that the
Langevin thermostat does not reproduce the correct hydrodynamics,
i.e., the hydrodynamic interactions are unphysically screened. In
order to correctly describe hydrodynamic interactions, one has to
resort to the DPD thermostat\cite{Soddemann:2003}.

In the past years DPD has established itself as a useful thermostat for
soft matter simulations.
The DPD thermostat is known to have several good properties, i.e., 
the thermostat satisfies Newton's third law by construction and 
owing to mass, momentum and temperature conservation, hydrodynamics is also
correctly reproduced \cite{Espanol:1995:2}. As it turns out, however, the DPD thermostat in its standard form 
is not capable of controlling liquid properties such as viscosity and diffusion constant\cite{Soddemann:2003}.
The aim of this work is to extend the DPD equations in such a way that these quantities
can be tuned by changing the parameters of the thermostat.
We consider the most general version of DPD\cite{Espanol:1995} 
and exploit the terms, which are not used in the standard DPD approach,
i.e., the damping of the perpendicular components of
the relative particle velocities.
This allows us to tune the viscosity of the coarse-grained liquid to
match that of the all-atom
counterpart while conserving the virtues of the standard DPD thermostat. 
This should be very useful for both coarse-grained 
and adaptive resolution simulations of soft matter. 

The article is organized as follows: In Sec. II the standard and 
new DPD thermostats are presented. The simulation setup is described in
Sec. III. The results of molecular dynamics (MD) simulations of the
Lennard-Jones fluid and liquid water are reported in Sec. IV, followed by the
conclusions in Sec. V.
\section{Thermostats}
\subsection{Standard DPD thermostat}
Newton's equations of motion are used in microcanonical
NVE MD simulations and generate dynamics with constant energy. 
In order to run simulations in the canonical NVT ensemble the equations of motion have
to be modified. In Langevin dynamics two additional forces are introduced, a~damping and
a~random force, whose ratio defines the temperature.
The DPD equations of motion (used by the DPD thermostat) are then given by\cite{Hoogerbrugge:1992,Espanol:1995}
\begin{equation}
\dot{\vec{r}}_i=\frac{\vec{p}_i}{m_i}~,
\end{equation}
and 
\begin{equation}
\dot{\vec{p}}_i=\vec{F}_i^\text{C}+\vec{F}_i^\text{D}+\vec{F}_i^\text{R}~,
\end{equation}
where $\vec{F}_i^\text{C}$ denotes the conservative force on the $i$th particle.
The damping and random forces, can be split up in particle pair
forces as
\begin{equation}
\vec{F}_i^\text{D}=\sum_{j\neq i}\vec{F}_{ij}^\text{D}~,
\end{equation}
\begin{equation}
\vec{F}_i^\text{R}=\sum_{j\neq i}\vec{F}_{ij}^\text{R}~,
\end{equation}
where the dissipative force reads as\cite{Espanol:1995}
\begin{equation}
\vec{F}_{ij}^\text{D}=-\zeta^\| w^\text{D}(r_{ij})(\hat{r}_{ij}\cdot\vec{v}_{ij})\hat{r}_{ij},
\label{F_D}
\end{equation}
and the random force is given by
\begin{equation}
\vec{F}_{ij}^\text{R}=\upsigma^\| w^\text{R}(r_{ij})\Theta_{ij}\hat{r}_{ij}~.
\label{F_R}
\end{equation}
In these equations the relative velocity $\vec{v}_{ij}=\vec{v}_i-\vec{v}_j$ between the $i$th and $j$th particle was  introduced, while $\hat{r}_{ij}$ denotes the unit vector 
of the interatomic axis $\vec{r}_{ij}=\vec{r}_i-\vec{r}_j$.
$\zeta^\|$ is the friction constant and $\upsigma^\|$ the noise strength.
$w^\text{D}(r_{ij})$ and $w^\text{R}(r_{ij})$ are $r$-dependent weight functions.
These are connected by the fluctuation-dissipation theorem (see Eqn.~\eqref{DFT1}).
The variable $\Theta_{ij}$ is symmetric in the particle indices ($\Theta_{ij}=\Theta_{ji}$) and has
the following first 
\begin{equation}
\langle\Theta_{ij}(t)\rangle=0~,
\label{theta_ij_1}
\end{equation}
and second moment
\begin{equation}
\langle\Theta_{ij}(t)\Theta_{kl}(t')\rangle=2(\delta_{ik}\delta_{jl}+\delta_{il}\delta_{jk})\delta(t-t')~.
\label{theta_ij_2}
\end{equation}
The fluctuation-dissipation theorem \cite{Espanol:1995} reads as
\begin{equation}
(\upsigma^\|)^2=\text{k}_\text{B}T\zeta^\|,
\label{DFT1}
\end{equation}
and
\begin{equation}
\left(w^\text{R}(r)\right)^2=w^\text{D}(r).
\label{DFT2}
\end{equation}
The above DPD equations conserve the total momentum and reproduce correctly
the hydrodynamics interactions in the system. However,
previous studies \cite{Soddemann:2003,Marsh:1997} have shown that the strength of the friction $\zeta^\|$ does not influence the
viscosity in linear order. In order to be able to tune the value of the viscosity of the
system while preserving all of the virtues of the standard
DPD thermostat presented briefly above we introduce in the next subsection its extension named
``Transverse DPD thermostat''.
\subsection{Transverse DPD thermostat}
We generalize Eqn.~\eqref{F_D} and Eqn.~\eqref{F_R} as
\begin{equation}
\vec{F}_{ij}^\text{D}=-\zeta w^\text{D}(r_{ij})\overleftrightarrow{P}_{ij}(\vec{r}_{ij})\vec{v}_{ij}
\end{equation}
and
\begin{equation}
\vec{F}_{ij}^\text{R}=\upsigma w^\text{R}(r_{ij})\overleftrightarrow{P}_{ij}(\vec{r}_{ij})\vec{\theta}_{ij}~,
\end{equation}
where $\zeta$  and $\upsigma$ are the friction constant and the noise
strength of the generalized thermostat, respectively (see the text below).
$\overleftrightarrow{P}_{ij}(\vec{r}_{ij})$ is a~projection operator
\begin{equation}
\overleftrightarrow{P}=\overleftrightarrow{P}^\text{T}=\overleftrightarrow{P}^2,
\end{equation}
which is symmetric in the particle indices ($\overleftrightarrow{P}_{ij}=\overleftrightarrow{P}_{ji}$).
The scalar noise (see Eqn.~\eqref{theta_ij_2}) is replaced by a~noise
vector $\vec{\theta}_{ij}$
\begin{equation}
\langle\vec{\theta}_{ij}(t)\otimes\vec{\theta}_{kl}(t')\rangle=2\overleftrightarrow{I}(\delta_{ik}\delta_{jl}-\delta_{il}\delta_{jk})\delta(t-t')~,
\end{equation}
which is antisymmetric in the particle indices ($\vec{\theta}_{ij}=-\vec{\theta}_{ji}$) due to 
the symmetry of the projection operator and the antisymmetry of the pair force (Newton's third law).
The corresponding Fokker-Planck operator $\mathcal{L}$ is a sum of two parts:
the deterministic part $\mathcal{L}_\text{D}$ 
\begin{equation}
\mathcal{L}_\text{D}
=
-\sum_i\left(
\frac{\partial}{\partial\vec{r}_i}\cdot\frac{\partial\mathcal{H}}{\partial\vec{p}_i}
-
\frac{\partial}{\partial\vec{p}_i}\cdot\frac{\partial\mathcal{H}}{\partial\vec{r}_i}
\right)
\end{equation} 
and the generalized DPD part $\mathcal{L}_\text{DPD}$
\begin{eqnarray}
\mathcal{L}_\text{DPD}
&=&
\nonumber
\zeta\sum_{i,j}w^\text{D}(r_{ij})
\frac{\partial}{\partial \vec{p}_i}
\cdot\left[
\overleftrightarrow{P}_{ij}
\left(
\frac{\partial\mathcal{H}}{\partial\vec{p}_i}-\frac{\partial\mathcal{H}}{\partial\vec{p}_j}
\right)
\right]
\\
\nonumber&&
+\frac{\upsigma^2}{2}\sum_{i,j}
w^\text{R}(r_{ij})^2
\frac{\partial}{\partial \vec{p}_i}\cdot
\left[\overleftrightarrow{P}_{ij}
\left(
\frac{\partial}{\partial \vec{p}_i}-\frac{\partial}{\partial \vec{p}_j}
\right)
\right].
\\
\end{eqnarray}
The equilibrium condition
$\mathcal{L}\text{e}^{-\beta\mathcal{H}}=
(\mathcal{L}_\text{D}+
\mathcal{L}_\text{DPD})\text{e}^{-\beta\mathcal{H}}=0$ then yields
the dissipation-fluctuation-theorem in the same form as given by Eqn.~\eqref{DFT1} and Eqn.~\eqref{DFT2}.

For the case where we choose the projector along the interatomic axis between particle $i$ and $j$
\begin{equation}
\overleftrightarrow{P}_{ij}(\vec{r}_{ij})=\hat{r}_{ij}\otimes\hat{r}_{ij},
\end{equation}
we retain the standard DPD thermostat.

Alternatively, one can project on the plane
perpendicular to the interatomic axis
\begin{equation}
\overleftrightarrow{P}_{ij}(\vec{r}_{ij})=\overleftrightarrow{I}-\hat{r}_{ij}\otimes\hat{r}_{ij}~.\label{pDPD}
\end{equation}
The space defined by the projector (\ref{pDPD}) is orthogonal to the case of the standard DPD and
introduces an extension of the  DPD thermostat, i.e., the Transverse DPD thermostat.
Note that owing to this orthogonality the new thermostat can be used in
combination with the standard one.
This enables us to adjust at the same time two friction constants $\zeta^\|$ and
$\zeta^\perp$ for the standard and Transverse DPD thermostats,
respectively. Galilei invariance remains valid by construction while
hydrodynamics is conserved within our error bar\cite{angular_momentum}.

Our basic assumption is that in contrast to the standard DPD the
viscosity is very sensitive to the damping perpendicular
to the interatomic axis. This damping mimics the shear of those DOFs that
were integrated out in the coarse-graining procedure.
In a system with two particles the stochastic forces of the
Transverse DPD thermostat act in the same direction as the shear forces.
The mean force acting on a particle in the sheared system with more
than two particles is hence a sum of two contributions: a force coming from the Transverse DPD thermostat
and another one originating from the shearing of the probe.
Therefore, the shear viscosity in a simulation with the Transverse DPD
thermostat is always higher than with the standard one.
In the Green-Kubo picture this additional viscosity arises 
from the projected velocity-velocity autocorrelation function, which is derived by 
the Mori-Zwanzig formalism \cite{Zwanzig:1960,Zwanzig:1961,Forster:1975}. The exact derivation is beyond the scope
of this paper and will be presented elsewhere.
Here, we shall demonstrate this by the results of
our numerical experiments presented in the results section.

\section{Simulation setup}

All simulations of the Lennard-Jones (LJ) liquid and liquid water are performed using
the ESPResSo package \cite{Espresso:2005}.

\subsection{Lennard-Jones Fluid}
We use the repulsive Weeks-Chandler-Andersen (WCA)
potential
\begin{equation}
U_\text{LJ}(r)=4\varepsilon\left(\frac{\sigma}{r^{12}}-\frac{\sigma}{r^6}+\frac{1}{4}\right)
\end{equation}
with the cutoff at $r_\text{c}=2^{1/6}\sigma$, $\sigma$ and $\varepsilon$
being the standard LJ parameters of length and energy.

We chose as the weight function (see Eqn.~\eqref{DFT2}) for both thermostats a~step function 
\begin{equation}
w^\text{D}(r)=
\left\{
\begin{array}{rr}
1,&r<r_\text{c}\\
0,&r\geq r_\text{c}
\end{array}
\right.
\end{equation}

The simulations are carried out with a~system consisting of 1000, 2000
and 4000 LJ particles at a temperature $T=1.2\varepsilon/k_B$ and density $\rho=N_\text{part}/V=1/(1.05\sigma)^3$ in
a~cubic box with periodic boundary conditions.

\subsection{Liquid Water}

All-atom water NVT simulations at ambient conditions are performed using the rigid
TIP3P water model\cite{Jorgensen:1983}. 
 The electrostatics is described by the reaction field (RF) method, in
 which all molecules outside a spherical cavity of a molecular based
cutoff radius $R_c=9$ \AA \; are treated as a dielectric continuum
with a dielectric constant $\epsilon_{RF}=80$~\cite{Neumann:1983,
Neumann:1985, Tironi:1995}. 
Typically, all-atom water simulations are carried out using global
thermostats, e.g., Berendsen\cite{Berendsen:1984}, 
Nos\'e-Hoover\cite{Nose:1984,Hoover:1985}, Nos\'e-Hoover
chains\cite{Martyna:1992} thermostats, that dissipate the energy uniformly in the system.
Local thermostats, e.g., Langevin,
DPD\cite{Soddemann:2003}, Andersen\cite{Andersen:1980},
Lowe-Andersen\cite{Lowe:1999, Peters:2004, Koopman:2006},
Nos\'e-Hoover-Lowe-Andersen\cite{Stoyanov:2005} thermostats, that dissipate energy on a spatially
localized scale are usually used in coarse-grained simulations.
Here, in order to reproduce the hydrodynamics
correctly we employed the DPD thermostat\cite{Soddemann:2003} with the
friction constant $\zeta^\|=0.5ps^{-1}$ and cutoff radius $R_c$. 
The constant $\zeta^\|$ is small compared to
the intrinsic friction coefficient $\xi$ of the TIP3P water system, i.e.,
$\xi=288.6ps^{-1}$, so that the stochastic dynamics 
yields the correct dynamics\cite{Kremer:1990,water2}.
For the coarse-grained water simulations with the Transverse DPD 
thermostat we have employed the single-site water model from Ref.\cite{Praprotnik:2007:2}, which reproduces essential
thermodynamics and structural properties, e.g., the pressure, density,
and radial distribution functions, of the all-atom rigid TIP3P water
model at standard conditions.
Other simulation details are the same as given in Ref.\cite{Praprotnik:2007:2}.
\section{Results}

\subsection{Lennard-Jones Fluid}

First, we checked in an 
equilibrium simulation the dependency of pressure and temperature on the strength of the friction.
We set the reference temperature to $1.2\varepsilon/k_B$ and measured the instantaneous
temperature defined as
\begin{equation}
T=\frac{2E_\text{kin}}{3N_\text{part}},\end{equation}
where $E_\text{kin}$ and $N_\text{part}$ are the kinetic energy and
the number of particles of the system, respectively.
The relative deviation between the measured and reference temperature was smaller then $1.2\%$ 
for all strengths of friction and all combination of thermostats.
The mean pressure at that temperature turned out to be $9.8\pm0.2 \varepsilon/\sigma^3$,
which is in a perfect agreement with the results of previous studies \cite{Dunweg:1993}.

Next, we studied the dependency of the liquid transport properties, i.e., the diffusion
constant and shear viscosity, on the friction constants
$\zeta^\|$ and $\zeta^\perp$ for the standard and Transverse DPD
thermostats, respectively.

\subsubsection{Diffusion constant}

\begin{figure}
\includegraphics[width=0.5\textwidth]{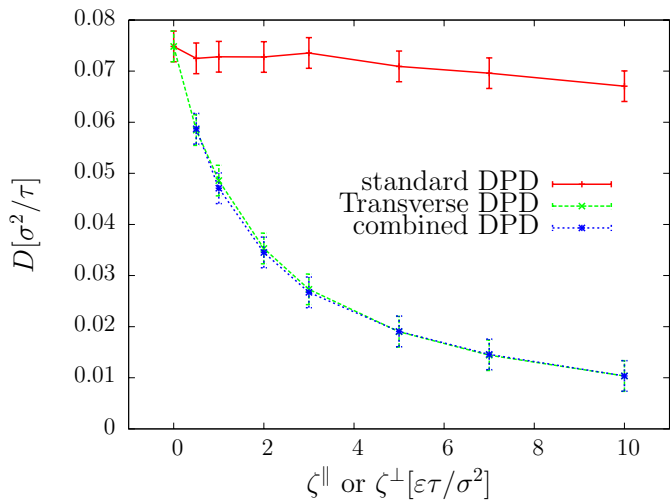}
\caption{\label{diff_eq} Diffusion constant (4000 LJ particles) as a
  function of the friction measured in equilibrium for different
  thermostats. In the case of the combined DPD thermostat only the strength of the
  friction parameter $\zeta^\perp$ was varied while the friction for the
  standard one was held constant at the value $\zeta^\|=1.0$.
The errors were obtained by averaging over several runs and Jackknife
analysis\cite{Efron:1982}.
}
\end{figure}

 The diffusion constant was computed from the particle displacements using the Einstein relation
\begin{equation}
\label{equ:diff}
D=\lim_{t\to\infty}\frac{|\vec{r}(t)-\vec{r}(0)|}{6t}.
\end{equation}
A small influence on this constant from the standard DPD thermostat
\cite{Soddemann:2003} is expected, but a~considerable one from the new Transverse
DPD thermostat.
Former results \cite{Soddemann:2003} for the standard DPD thermostat could be
confirmed. The value of the diffusion constant approaches the
equilibrium value ($D=0.08 \sigma^2/\tau$) for vanishing friction of the
Transverse DPD thermostat (see Fig. \ref{diff_eq}).
The diffusion constant $D$ is very sensitive on the 
friction $\zeta^\perp$ for the Transverse DPD thermostat. By changing
$\zeta^\perp$ it is therefore possible to tune the diffusion constant.

\subsubsection{Shear viscosity}
The viscosity was measured in nonequilibrium molecular dynamics (NEMD)
simulation by shearing the system with a constant shear rate in the $y$-direction\cite{Soddemann:2003}
\begin{equation}
\dot{\gamma}=\frac{\partial u_x}{\partial y}~.
\end{equation}
The viscosity can then be determined by the following simple formula
\begin{equation}
\eta=\frac{F}{\dot{\gamma}L^2}~,
\end{equation}
where $F$ is the mean force (momentum transfer per time unit).
The apparent shear viscosity $\eta$ measured in NEMD simulation approaches 
the equilibrium viscosity 
with decreasing shear rate.

\begin{figure}
\includegraphics[width=0.5\textwidth]{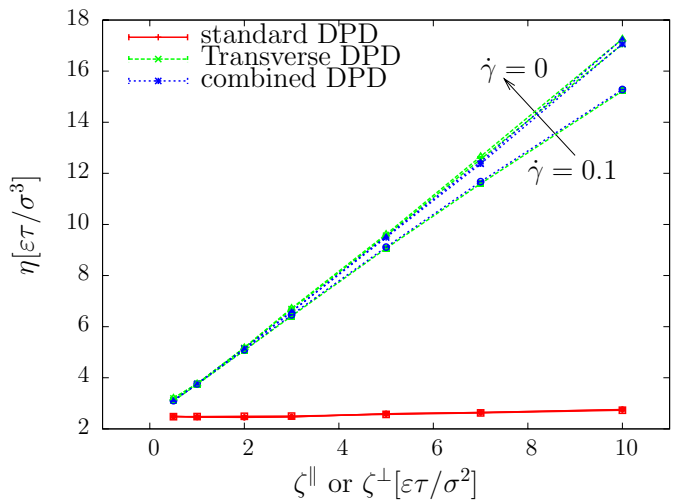}
\caption{\label{nemd_visco} Shear viscosity measured with the NEMD
  algorithm for different thermostats (4000 LJ particles) and the different shear rates (0.1 and 0.01 shown),
  which then are extrapolated to vanishing shear rate.
The errors are obtained by Jackknife analysis \cite{Efron:1982}.}
\end{figure}

We found nearly no dependency on the strength of the friction
$\zeta^\|$ for the standard DPD thermostat, (see also Ref. \cite{Soddemann:2003}). 
In contrast the friction $\zeta^\perp$ for the Transverse DPD
thermostat gives a~very sensitive means of controlling the viscosity (see Fig.~\ref{nemd_visco}).
In the case of the combination of both thermostats the shear viscosity
is mostly controlled by the Transverse DPD thermostat.
In the limit of a vanishing shear rate a value of $2.45\pm0.07 \varepsilon\tau/\sigma^3$ 
was extrapolated, which matches former results \cite{Dunweg:1993}. 
This extrapolation has to be done due to short characteristic timescale of the LJ system.
Additionally, we also checked that the equilibrium correlation of the
pressure tensor is in accordance with the Green-Kubo
picture: all (non-auto) off-diagonal - (off-)diagonal 
elements are uncorrelated, (60 of 81 possible elements are zero).

For higher values of $\zeta^\perp$ the apparent viscosity becomes increasingly
dependent on the shear rate. This is due to the fact
that the dynamics is controlled in this regime by the thermostat forces
(that are linear in $\zeta^\perp$) and hence we end up measuring the ``viscosity of the thermostat''.

\subsection{Tuning the Dynamics of Water}

Having shown that the new thermostat enables us to tune the
diffusion constant and viscosity of a simple liquid we apply
it to an important physical example, i.e., liquid water at ambient conditions.

\begin{figure}
\includegraphics[width=0.5\textwidth]{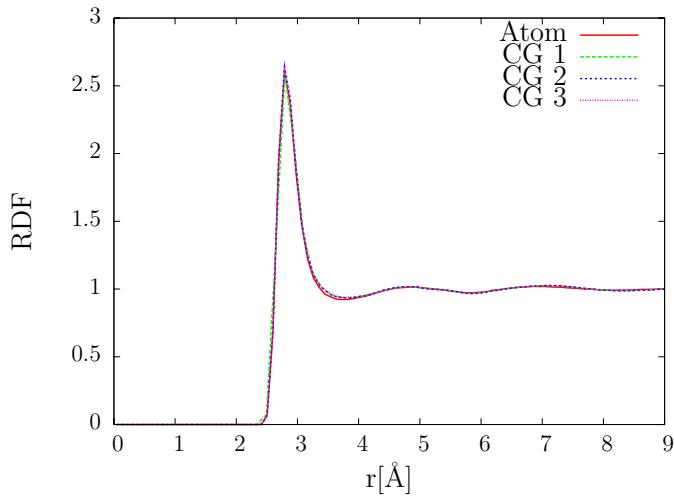}
\caption{\label{water_rdf} The center-of-mass 
radial distribution functions of the all-atom ($\zeta^\|=0.5ps^{-1}$,
$\zeta^\perp=0$) and 
several coarse-grained simulations ($\zeta^\|=0$ and $\zeta^\perp_1=0.5ps^{-1}$, $\zeta^\perp_2=0.75ps^{-1}$ and $\zeta^\perp_3=1.0ps^{-1}$).}
\end{figure}

We first check that the structural properties do not depend on the
thermostat and also that we obtain consistency between 
the coarse-grained and atomistic simulations.
The center-of-mass radial distribution functions of the all-atom and
coarse-grained system using different values of $\zeta^\perp$ match
within the line thickness (see Fig.~\ref{water_rdf}).

There is an intrinsic timescale difference in the diffusive dynamics
of the coarse-grained water system because of the reduced number of
DOFs, i.e., the self-diffusion constant for the
coarse-grained water model is approximately $2$ times larger then the corresponding
all-atom one using the Langevin thermostat with the same background
friction in both cases\cite{Praprotnik:2007:2}.
We used $\zeta^\perp=0.8ps^{-1}$ for the Transverse DPD thermostat ($\zeta^\|=0$)
to match the diffusion constant of the coarse-grained water model to
the corresponding value $D=3.0\cdot10^{-9}\,\text{m}^2/\text{s}$
obtained from the all-atom simulation with the standard DPD thermostat.
With a friction strength of $\zeta^\perp=0.6ps^{-1}$, which is as
desired very close to the above value for matching the diffusion constants, 
we were also able to match the viscosity 
to the desired value $\eta=0.5\pm 0.1\cdot 10 ^{-3}\text{Pa}\cdot\text{s}^{-1}$ for the TIP3P water model 
(from our atomistic simulation with $\zeta^\|=0.5ps^{-1}$).
In this case the characteristic times for the atomistic model are much
longer than the time scale of the shearing ($1/\dot{\gamma}$).
Therefore even with a shear rate of $\dot{\gamma}=0.01$ we are in
the no shear limit and hence no extrapolation is required.
The obtained diffusion constants and viscosities are in good
agreement with the published data\cite{Wu:2006}.

\begin{figure}
\includegraphics[width=0.5\textwidth]{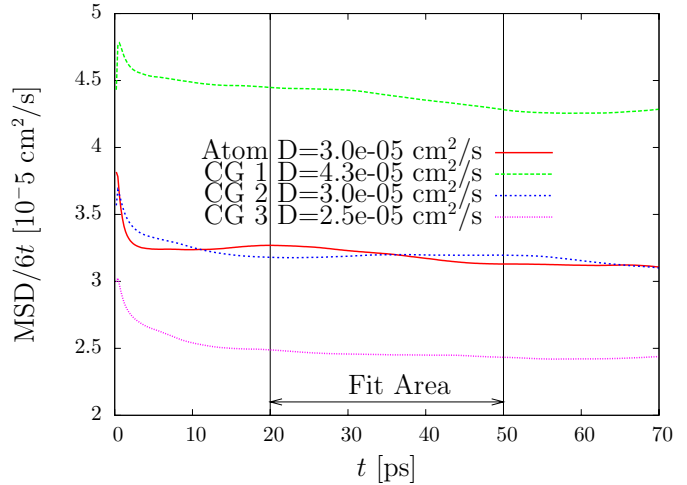}
\caption{\label{water_diff} The mean square displacements over time
  plot of the all-atom ($\zeta^\|=0.5ps^{-1}$,  $\zeta^\perp=0$) and
several coarse-grained simulations ($\zeta^\|=0$ and $\zeta^\perp_1=0.5ps^{-1}$, $\zeta^\perp_2=0.875ps^{-1}$ and $\zeta^\perp_3=1.0ps^{-1}$).}
\end{figure}

Thus, employing the new thermostat one can reproduce both the
structure and the dynamics of the all-atom
liquid water with the single-site coarse-grained
water model. This is essential for synchronizing the timescales of the all-atom and coarse-grained regimes in the
adaptive resolution MD simulations\cite{Praprotnik:2007:2}. 

\section{Conclusions}
In this paper we introduced an extension of the DPD
thermostat that allows for controlling the transport properties of molecular liquids, e.g., water, while
preserving the hydrodynamics.

The presented Galilean invariant local thermostat can be used in coarse-grained simulations to tune
the diffusion constant and viscosity of the system to the desired values.
This opens up the possibility of reproducing the atomistic dynamics by coarse-grained simulations,
as it is required, for example, in recently introduced adaptive resolution simulations.
\subsection*{Acknowledgements}
We are grateful to R.~Delgado-Buscalioni and B.~D{\"u}nweg for
discussions at early stage of this work. We also thank J. Kirkpatrick
for critical reading of the manuscript.

%


\begin{thebibliography}{10}
\bibitem{angular_momentum}
Note that due to the Transverse DPD thermostat forces between molecules are not
  central and therefore the local angular momentum is not
  conserved. Nevertheless, simulation results show that these angular
  momentum fluctuations cancel out on average and hence the total angular
  momentum of the system is almost a conserved quantity.
\end{thebibliography}

\end{document}